\documentclass[amsmath,twocolumn,superscriptaddress,prl,aps]{revtex4}
\usepackage{amsthm,amsfonts,graphicx,verbatim}
\newcommand{\be}{\begin{equation}}
\newcommand{\ee}{\end{equation}}
\newcommand{\bea}{\begin{eqnarray}}
\newcommand{\eea}{\end{eqnarray}}

\renewcommand{\phi}{\varphi}
\renewcommand{\epsilon}{\varepsilon}
\renewcommand{\vec}[1]{{\bf #1}}

\renewcommand{\Re}{{\rm Re}\,}

\renewcommand{\cite}[1]{[\onlinecite{#1}]}

\begin{document}

\title{Prediction and description of a chiral pseudogap phase}
% \author{people}
% \affiliation{places}
\author{Rahul Nandkishore}
\affiliation{Department of Physics, Massachusetts Institute of Technology, Cambridge MA 02139, USA}
%\author{Liang Fu}
%\affiliation{Department of Physics, Massachusetts Institute of Technology, Cambridge MA 02139, USA}

%\date{\today}

\begin{abstract}
We point out that a system which supports chiral superconductivity should also support a chiral pseudogap phase: a finite temperature phase wherein superconductivity is lost but time reversal symmetry is still broken. This chiral pseudogap phase can be viewed as a state with phase incoherent Cooper pairs of a definite angular momentum. This physical picture suggests that the chiral pseudogap phase should have definite magnetization, should exhibit a (non-quantized) charge Hall effect, and should possess protected edge states that lead to a quantized thermal Hall response. We explain how these phenomena are realized in a Ginzburg-Landau description, and comment on the experimental signatures of the chiral pseudogap phase. We expect this work to be relevant for all systems that exhibit chiral superconductivity, including doped graphene and strontium ruthenate.% Moreover, since the chiral pseudogap phase can survive higher temperatures and higher levels of disorder than the chiral superconductor, we expect it to have strong advantages for nano
\end{abstract}

\maketitle

Chiral superconductors feature pairing gaps that wind in phase around the Fermi surface (FS), breaking time reversal symmetry (TRS) \cite{Volovikold, Sigrist, Sigrist_1, Vojta}. They realize topological superconductivity \cite{Kane, Qi, Cheng, Schnyder} and exhibit a host of fascinating and technologically useful properties, such as protected edge states, a quantized thermal Hall effect, and unconventional zero modes in vortices \cite{Volovikold, Sigrist, Sigrist_1, Vojta}. Chiral p-wave superconductivity is believed to have been found in strontium ruthenate \cite{Mackenzie}, and chiral $d$ wave superconductivity has been established to be the leading weak coupling instability in strongly doped graphene \cite{Nandkishore, Thomale}. However, all theoretical work to date has focused on chiral superconductors at low temperatures. In this work, we argue that much of the exotic phenomenology associated with chiral superconductivity can be exhibited even at high temperatures, when the superconductivity is absent. In particular, we predict the existence of a chiral pseudogap phase with a rich phenomenology, including a magnetic dipole moment, non-quantized charge Hall effect, protected edge states, and quantized thermal Hall effect. Since this pseudogap phase does not require low temperatures or exceptionally clean systems (unlike chiral superconductivity), we expect it to be advantageous for nanoscience applications. 

Emery and Kivelson have pointed out that phase incoherent Cooper pairs can form at temperatures much higher than the characteristic temparature for onset of superconductivity \cite{Kivelson}. This `preformed Cooper pairs' picture has been invoked as a possible explanation for the pseudogap phase of the cuprate high-$T_c$ materials. The nature of the pseudogap phase of the cuprates remains controversial, not least because the pre-formed Cooper pairs picture for the cuprates does not lead to many sharp testable predictions. However, the situation is markedly different for a chiral superconductor, where a pseudo-gap phase with phase incoherent pre-formed Cooper pairs can still break time reversal symmetry. The resulting chiral pseudo-gap phase has a rich and distinctive phenomenology, which should be readily testable experimentally. 

In this letter, we provide a Ginzburg-Landau description of the physics of a chiral superconductor that both demonstrates that a chiral pseudo-gap can exist, and also elucidates the phenomenology of the chiral pseudogap phase. We work for simplicity in the continuum, assuming continuous rotation symmetry, although we expect our results to apply also for lattice systems. We work in two spatial dimensions, since most materials where chiral superconductivity is expected to arise are either two dimensional, or layered three dimensional materials. We comment in the end on the likely implications of our results for systems that are expected to exhibit chiral superconductivity, such as strontium ruthenate and doped graphene. 

\begin{figure}
\includegraphics[width = \columnwidth]{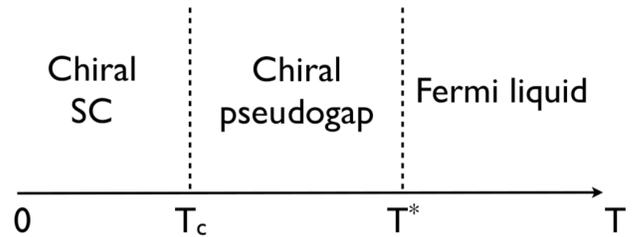}
\caption{\label{fig: fig1} Schematic phase diagram of material exhibiting chiral superconductivity. At high temperatures, the system is a Fermi liquid. At low temperature, it is a chiral superconductor. At intermediate temperatures, there arises a chiral psedudogap phase, marked by phase incoherent Cooper pairs of definite angular momentum. The chiral pseudogap phase breaks time reversal symmetry, has a definite magnetization, and inherits the topological structure of the chiral superconductor, possessing protected edge states that lead to a quantized thermal Hall response. In addition, the chiral pseudogap also exhibits a (non-quantized) DC charge Hall effect, which is forbidden in the chiral superconductor.}
\end{figure}

The broad picture is summarized in Fig.1. At the lowest temperatures $T<T_c$, the system is a chiral superconductor. At high temperatures $T>T_*$, the system is a Fermi liquid. At intermediate temperatures, $T_c<T<T_*$, there arises a chiral pseudogap phase with phase incoherent Cooper pairs of a definite angular momentum (Fig. 2). This chiral Cooper pairing produces a magnetization which may be detected using torque magnetometry. Moreover, the chiral pseudogap phase inherits the topological properties of the chiral superconductor, and in particular has protected edge states that lead to a quantized thermal Hall response. Finally, at finite temperature the chiral pseudogap phase also has a (non-quantized) bulk charge Hall response, even in the clean, DC limit. Such a response is forbidden in the chiral superconductor (which cannot support a voltage gradient), and thus the phenomenology of the chiral pseudo-gap state is in some ways even richer than that of the chiral superconductor. 

We note that a system with a $U(1)\times Z_2$ symmetry, and an intermediate phase with broken $Z_2$ symmetry only, was also discussed in \cite{Xu}. However, the physics of these phases is very different, since in the case \cite{Xu}, the $Z_2$ symmetry was not associated with time reversal, and nor was the system coupled to an electromagnetic gauge field. 

%The central observation is that although the chiral superconductor breaks time reversal symmetry in addition to a U(1) gauge symmetry, these two symmetries need not be broken together. In particular, at finite temperature, phase coherence can be lost due to thermal fluctuations while the time reversal symmetry is still broken. The resulting chiral pseudo-gap phase

{\it Ginzburg-Landau theory:}
We consider a two dimensional electron system with continuous rotation symmetry, and with a doubly degenerate Cooper instability. Cooper pairing occurs in a channel with angular momentum $l$, and the Cooper pair wavefunctions that vary around the Fermi surface as $\psi_1 = \eta_1 \cos l \phi$ and $\psi_2 = \eta_2 \sin l \phi$ respectively are assumed to have the same energy, where the angle $\phi$ parametrizes position on the two dimensional Fermi surface. The degeneracy of the two pairing functions follows from rotation symmetry. An analogous situation is believed to arise in strontium ruthenate (with $l=1$) \cite{Mackenzie} and also in doped graphene (with $l=2$) \cite{Nandkishore}. We focus on the spin-singlet case for simpicity, although we expect our results to apply also to spin triplet systems. The gap function behaves as 
\begin{equation}
\Delta(\vec{r},t, \phi) = \eta_1(\vec{r},t) \cos l \phi + \eta_2(\vec{r},t) \sin l \phi
\end{equation}
We wish to construct a Ginzburg-Landau free energy functional out of these gap wavefunctions. The free energy functional should contain all symmetry allowed terms. Fortunately, the restriction to a system with continuous rotation symmetry greatly restricts the allowed terms in the Free energy functional. We also assume that the Free energy functional should respect time reversal symmetry and inversion symmetry.
%We consider pairing with angular momentum $\pm 2$, as the simplest chiral spin-singlet state. The $d_{x^2-y^2}$ and 
To quartic order in the gap functions, the static part of the free energy, $F_0$, takes the form
\begin{equation}
F_0 = (T-T_{MF}) (|\eta_1|^2 + |\eta_2|^2) + K_1 (|\eta_1|^2 + |\eta_2|^2)^2 + K_2 |\eta_1^2 + \eta_2^2|^2
\end{equation}
where $T_{MF}$ is the temperature for onset of Cooper pairing, and $K_{1,2}>0$. %In terminating the expansion at quartic order in gap functions, we have assumed that we are working at a temperature close to $T_{MF}$. 
The positivity of $K_2$ meains that this system favors co-existence of the two orders with relative phase $\pm i$. As a result, the order parameter takes the form $\Delta (\vec{r},t,\phi)= \Delta(\vec{r},t) \exp(\pm i l \phi)$.%, where $\phi_k$ is the angle around the Fermi surface.  

To obtain the physics we are interested in, we must examine the gradient terms in the Free energy. Since we are interested in finite temperature physics, we neglect temporal fluctuations, and consider a time independent Ginzburg-Landau description of the problem, with $\eta_{1,2}(\vec{r},t) = \eta_{1,2}(\vec{r})$. Note that since we are dealing with a charged system, we must take into account the coupling to the electromagnetic field. The electromagnetic field is introduced by minimal coupling, through $i \partial \rightarrow i D = i \partial + 2 e A$ (the factor of $2$ arises because we have Cooper pairs). We define $\vec{D}$ to be a two component vector, $\vec{D} = (i\partial_x + 2 e A_x, i\partial_y + 2 e A_y)$. At second order in gradients, the Free energy has the form
\begin{widetext}
\begin{eqnarray}
F_{\nabla} &=& J_1(\vec{D} \eta_1)^* \cdot \vec{D}\eta_1 +  J_1(\vec{D} \eta_2)^* \cdot \vec{D} \eta_2 + J_2 (\vec{D} \eta_1)^* \times \vec{D} \eta_2\cdot \vec{ \hat z}- J_2 (\vec{D} \eta_2)^* \times \vec{D} \eta_1 \cdot \vec{\hat z} \\
&=&  J_1 (\vec{D} \eta_1)^* \cdot \vec{D}\eta_1 + J_1  (\vec{D} \eta_2)^* \cdot \vec{D} \eta_2 +  J_2 \eta_1^* (\vec{D}^* \times \vec{D})\cdot \vec{\hat z} \eta_2 -  J_2 \eta_2^* (\vec{D}^*\times \vec{D}) \cdot \vec{\hat z} \eta_1
\end{eqnarray}
\end{widetext}
where $J_{1,2}$ are phenomenological constants, and in going from the first line to the second we have performed an integration by parts and have ignored possible boundary terms. %We have assumed for simplicity that temporal gradients cost the same amount of energy as spatial gradients, but this assumption may be relaxed without altering our results. 
We have used the notation $(\vec{D}^*\times \vec{D})\cdot \vec{\hat z} = [(-i \vec{\partial} + 2 e \vec{A}) \times (i\vec{\partial} + 2 e \vec{A})] \cdot \vec{\hat z} = \partial \times \partial -  2 e i (\partial_x A_y - \partial_y A_x)=  \partial \times \partial - 2 e i B_z$, where $B_z$ is the magnetic field transverse to the plane \cite{Mineev}. 

We now write $\eta_{1,2} = \Delta_0 \exp(i\theta_{1,2})$, and we neglect (massive) fluctuations in the magnitude $\Delta_0$. We define new variables $\theta_{+} = \frac12 (\theta_1 + \theta_2)$ and $\theta_- = \theta_1-\theta_2$, which are $2\pi$ periodic (a $2\pi$ shift in either leaves the physical state unchanged). Recalling that $\partial \times \partial \theta_{1,2} = n_v$, where $n_v$ is the vorticity, we find that the free energy takes the simple form $ F = F_0(|\Delta_0|)+ |\Delta_0|^2 (F_+ + F_- + F_{+-})$, where 
\begin{equation}
F_+ = 2  J_1 \left[ (i\vec{\partial} \theta_+ + e \vec{A}) \cdot (i\vec{\partial} \theta_+ + e \vec{A}) \right]\label{eq: fplus}
\end{equation}
\begin{equation}
F_-= \frac12  \left[J_1 (\partial \theta_-)^2 + 8 J_2 B_z \sin \theta_- \right] + 2 K_2 |\Delta_0|^2 \cos 2 \theta_- \label{eq: fminus}
\end{equation}
\begin{equation}
F_{+-} = 2 J_2 \sin \theta_- n_v
\end{equation}
%
%Thus we find that at quadratic order in gradients, the $\theta_+$ and $\theta_-$ sectors decouple. (Couplings arising from higher order gradient terms should be irrelevant in the renormalization group sense and may be ignored). 

We now interpret these equations. Firstly, note that the anisotropy term in (\ref{eq: fminus}) favors $\sin \theta_- = \pm1$. Next, note that $\sin \theta_-$ couples to an external magnetic field in the same way as a magnetization, and can be interpreted as the magnetization associated with the orbital angular momentum of the chiral Cooper pairs. We point out that $\sin \theta_-$ serves as an Ising order parameter for time reversal symmetry breaking, and couples to the vortex density (which is expected since the vortices carry magnetic flux). We also note that the $\theta_-$ sector (at $ B_z = 0$) describes an XY model with Ising anisotropy. The Ising anisotropy is known to be a relevant perturbation \cite{JKNN} and thus at low enough temperatures the phase $\theta_-$ will be pinned to $\sin \theta_- = \pm 1$, for arbitrarily weak $K_2$. The $\theta_-$ order will be destroyed by thermal fluctuations at a temperature $T_*$, above which the time reversal symmetry will be restored. 

%A fourfold anisotropy is known to be marginally irrelevant at the XY critical point, however, for sufficiently large $K_2$, the fourfold anisotropy becomes relevant, and pins the $\theta_-$ field \cite{Sachdev}. We assume that $K_2$ is large enough to be relevant, such that at a critical temperature $T^*$ there is a transition from a state with $\sin \theta_- = \pm 1$ to a state where $\theta_-$ is disordered. 
%The %While $\theta_-= \pi/2, -3\pi/2$ are physically equivalent, this Dirac string will nonetheless have some line tension, which will tend to bind the $2\pi$ vortices into $4\pi$ vortices. If $\theta_-$ is truly pinned, the 
%disordering transition in the $\theta_+$ sector will thus have to proceed via proliferation of $hc/e$ vortices {\bf [XY*?]}. 
% suggests that the chiral pseudogap phase should have a magnetic dipole moment, arising from the orbital angular momentum of the Cooper pairs. This is confirmed by Eq.\ref{eq: fminus}, which shows that the Ising order parameter $\sin \theta_- = \pm 1$ couples directly to an external magnetic field. Indeed, we can identify from Eq.\ref{eq: fminus} that the magnetization is directly proportional to the Ising order parameter $\sin \theta_-$. The magnetization of the chiral pseudogap phase should be easily measurable via torque magnetometry. 

Meanwhile, the $\theta_+$ sector is the action of a superconductor. It will have a Higgs phase, where the photon is massive, and $\theta_+$ is locked, and it will also have a trivial phase, where $\theta_+$ is disordered, with the phase transition occurring at a temperature $T_c$. The disordering transition will be associated with proliferation of vortices in $\theta_+$. The magnetization $\sin \theta_-$ imbalances the number of vortices and anti-vortices, so the vortex proliferation transition falls into the universality class of XY models in a magnetic field studied in \cite{Sondhi}, and not the usual KT universality class.

Finally, we note that the elementary vortices in the theory carry magnetic flux $hc/e$ (instead of the more usual $hc/2e$), and involve a simultaneous $2\pi$ winding in $\theta_2$ and $\theta_2$. This is because a vortex in $\theta_1$ or $\theta_2$ alone leaves the phase $\theta_-$ misaligned everywhere around it, and hence carries a large anisotropy energy cost. As a result, the loss of superconductivity at $T_c$ is associated with a proliferation of double vortices, carrying magnetic flux $hc/e$, in an external magnetic field.

Now the temperatures $T_c$ and $T^*$ are in principle independent. We assume that $T^* > T_c$ for the purposes of this paper. Since $Z_2$ symmetry breaking is more robust against thermal fluctuations than $U(1)$ symmetry breaking, we believe $T^* > T_c$ represents the generic case for a chiral superconductor. %the ordered state for $\theta_-$ has no gapless fluctuation modes, we expect the effect of thermal fluctuations to be weaker in $\theta_-$ sector, and we therefore expect the $\theta_-$ order to survive to higher temperatures than the $\theta_+$ order. Thus, we believe $T^* > T_c$ represents the generic case for a chiral superconductor. 

 %The Ising anisotropy is known to be relevant \cite{Sachdev}, thus this system falls into the universality class of the Ising model. It will have an ordered phase, where $\sin \theta_-$ is pinned at $\pm 1$, and a disordered phase. Note in particular that the ordering temperature for the Ising transition is independent of the ordering temperature for the superconducting sector. Since the Ising ordered state has no gapless fluctuation modes, we expect the effect of thermal fluctuations to be weaker in the Ising sector, and we therefore expect the Ising order to survive to higher temperatures than the superconducting phase coherence. This leads to our prediction of a chiral pseudogap phase (Fig.1,2). 

Now, since the chiral pseudogap state gaps out the full Fermi surface, there should be a gap to quasiparticle excitations. However, since the chiral pseudogap state only forms at finite temperature, there will always be some thermal excitation across the gap. These thermally excited quasiparticles will see the magnetic field from the chiral Cooper pairing, and will give rise to a (classical) charge Hall effect (and thermal Hall effect), in complete analogy with Hall effect in an external magnetic field. Note that the chiral pseudo-gap phase displays a bulk charge Hall effect even in the DC limit, with a clean sample, whereas the chiral superconductor cannot display a Hall conductance in this limit \cite{Roy}, because a superconductor cannot sustain a voltage drop. Similar effects should also arise from the effect of the magnetic field on the charged Cooper pairs. However, none of these effects will be quantized, or topological.

\begin{figure}
\includegraphics[width = \columnwidth]{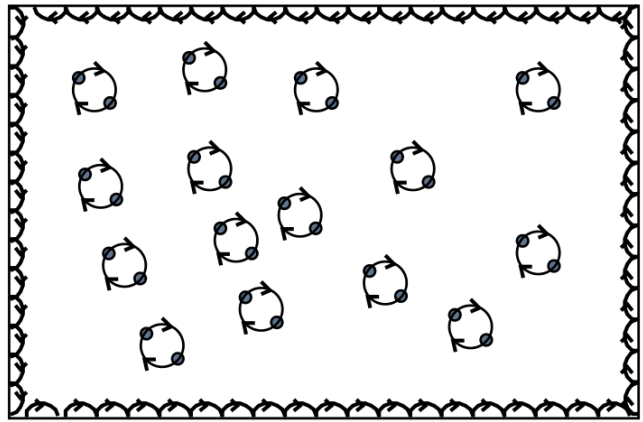}
\caption{\label{fig: fig2} The chiral pseudogap phase has chiral, phase incoherent Cooper pairs in the bulk, which have a definite angular momentum. This picture suggests that there should be edge states associated with skipping orbits for the Cooper pairs, as shown above. This intuition manifests itself mathematically through a Chern number for the quasiparticle excitations of the chiral pseudogap phase. The orbiting Cooper pairs also produce a magnetization, which couples directly to an external magnetic field, and also to vortices.}
\end{figure}

{\it Topological properties and signatures in transport:}

The physical picture of the chiral pseudogap state Fig.2 suggests there should be edge modes associated with `skipping orbits' for the Cooper pairs, which should also give rise to some kind of quantized Hall effect. A good way to understand this is as follows: the (fermionic) quasiparticle spectrum of the chiral superconductor is gapped, and consists of bands that carry Chern number. This is the origin of the quantized thermal Hall response of the chiral superconductor \cite{Senthil}. Crucially, the sign of the Chern number depends only on the sign of the time reversal symmetry breaking (i.e. whether $\theta_- = \pm \pi/2 \quad {\rm mod} \quad 2 \pi$), and is independent of $\theta_+$. When we disorder $\theta_+$ to go from the chiral superconductor into the chiral pseudogap phase, the fermion spectrum remains gapped, and continues to carry a Chern number. Thus, the edge states and quantized thermal Hall response exhibited by the chiral superconductor should also be exhibited by the chiral pseudogap phase.

While the thermal Hall conductance coming from the edge states is quantized, experimentally, one measures the total Hall conductance, obtained by summing over topological and quasiparticle contributions, and the contribution from thermally excited quasiparticles is not quantized. However, we could imagine isolating the `topological' Hall effect from the classical Hall effect if the width of the pseudogap region is sufficiently large. The gap to quasiparticle excitations is of order $T_*$, whereas the topological Hall conductance appears already at $T>T_c$. Therefore, if $T_c\ll T_*$ (i.e. if the chiral pseudogap region is sufficiently wide), then the fermionic excitations can be supressed arbitrarily strongly by working deep in the pseudogap regime, at $T_c<T\ll T_*.$ The theoretically clean limit involves taking $T_c \rightarrow 0$, whereupon there will be a truly quantized thermal Hall response. %Moreover, in the limit $T_c \rightarrow 0$ one can also study the quantum phase transition from the chiral superconductor to the chiral pseudogap state. If $T_c = 0$ for the $d+id$ state, and one destroys the superconductivity by proliferating $4\pi$ vortices in $\theta_+$, then the chiral pseudogap state that one obtains is identical to the chiral spin liquid \cite{?}. This has a spin quantum Hall response in addition to a thermal quantum Hall response, which originates because the quasiparticles also carry a spin quantum number. 

If the quasiparticles of the chiral pseudogap state carry any quantum numbers, the chiral pseudogap state will display the corresponding quantized Hall effects in addition to the quantized thermal Hall effect. For example, the quasiparticles of the $d+id$ pairing state carry spin, and the $d+id$ superconductor displays a spin quantum Hall effect \cite{Senthil}. The chiral pseudogap phase produced by thermally disordering the $d+id$ superconductor should inherit this topological response, and should display a spin quantum Hall effect in addition to a thermal quantum Hall effect. 

A provocative question is whether it is possible for the chiral pseudogap state to have a charge quantum Hall response -i.e. whether it is possible to have a direct second order transition from a chiral superconductor to a quantum anomalous (charge) Hall state. While this is clearly not the generic case, there seems to be no reason why such a transition should be impossible, and we believe it is a fruitful topic for further investigation. 

{\it Optical signatures:} Since the chiral pseudogap phase has a Hall conductivity (at finite temperature), it should exhibit a Kerr effect (rotation of the polarization angle of reflected light). It was demonstrated in \cite{Kerr} that the Kerr rotation arising from reflection from a single two dimensional sheet of Hall conductance $\sigma_{xy}$ placed on a substrate of dielectric constant $n$ is 
\begin{equation}
\theta_K \approx \frac{8 \pi \Re \sigma_{xy}}{c(n^2-1)}% \approx \frac{4 \alpha l \sin \theta_- }{n^2-1}\label{eq: kerr}
\end{equation}
The Kerr signal arising due to a charge Hall response should be much stronger than the Kerr signal in the absence of a charge Hall response, and thus should be easier to detect experimentally. %Where $\alpha = 1/137$ is the fine structure constant, we have assumed that $k=l$, and have also assumed that the experiment is done at frequencies much smaller than the gap scale $\omega \ll \Delta$, so that the Hall conductivity is given approximately by its DC value. This is a signal of order $10^{-2}$ radians, which is much larger than the signals of order $10^{-9}$ radians which have been measured using Sagnac interferometry \cite{Xia}, and as such it should lie well within the reach of existing experimental techniques. 
One can also consider a layered system with multiple layers of chiral pseudogap phase. It is simplest to consider the limit when there is no tunneling of quasiparticles or Cooper pairs between layers, so that the different layers are coupled only through the electromagnetic field. In this situation, the magnetic dipole interactions between layers should align the layers so that the same sense of chiral pseudogap phase is obtained in each layer i.e. $\sin \theta_- = +1$ in all layers, or else $\sin \theta_- = -1$ in all layers. The Kerr response from this system is analogous to the Kerr response of a ferromagnet, discussed in \cite{Argyres}

{\it Domain structure:} We note that we have this far assumed that the Ising sector ($\sin \theta_-$) has long range order. However, an Ising system can have domains, and the chiral pseudogap state will indeed have domains with $\sin \theta_- = \pm 1$. In the absence of magnetic disorder, the domain size will be determined by the competition between the gradient energy cost of domain formation, and the long range magnetic dipole interactions between domains, in analogy with the electronic micro-emulsion phases from \cite{Kivelson2}. If the characteristic domain size is larger than the system size (or the laser spot size for Kerr effect measurements), then the domain structure can be neglected, otherwise it will suppress the experimental signal by a factor of $\sqrt{N}$, where $N$ is the number of domains. 

{\it Relevance for experimental systems:} The only material which is known to be a chiral superconductor is strontium ruthenate. We expect that a chiral pseudogap phase will form in this material, although it will be somewhat more complicated than the chiral pseudogap phase discussed above because the pairing in strontium ruthenate is spin triplet. Meanwhile, graphene doped to $3/8$ or $5/8$ filling is expected to exhibit spin singlet, d wave chiral superconductivity if interactions are sufficiently weak \cite{Nandkishore, Thomale}. Here too there should be a chiral pseudogap phase, without any of the complications arising from spin triplet pairing. 

The separation of scales between $T_*$ and $T_c$ is largest for systems with small carrier densities \cite{Kivelson}. For strongly doped graphene, the carrier density will be very high, to $T_c$ is likely to be quite close to $T_*$. In this case, the chiral pseudogap phase will not be very relevant. However, if the carrier density is strongly depleted, for example by localization on disorder, then the separation between scales might become more substantial. Since all known methods of doping graphene to the Van Hove point introduce large amounts of disorder, it remains possible that the chiral pseudogap phase might be discovered in doped graphene. 

Of course, the best studied superconducting materials are probably the cuprates. While we are not aware of any compelling scenario for chiral superconductivity (or chiral pseudogaps) in the cuprates, it is amusing to note that there are numerous experiments suggesting both time reversal symmetry breaking and a net vorticity in the pseudogap phase of the cuprates \cite{Xia, Ong}. A more detailed analysis of these experiments is beyond the scope of the present work. 

{\it Outlook}: The paucity of known chiral superconductors makes it difficult to find materials that exhibit chiral pseudogaps. However, given the intense current interest in searching for realization of chiral superconducitivity, it seems inevitable that such materials will eventually be found. At that point, a chiral pseudogap phase will likely be found also. Moreover, we know from the existing high-$T_c$ materials that pseudogap phases can survive to much higher temperatures than superconductivity. Since the chiral pseudogap phase exhibits much of the exotic phenomenolgy of the chiral superconductor, and may survive to significantly higher temperatures, we expect that the chiral pseudogap phase will not only be of intrinsic theoretical interest, but will also be highly beneficial for nanoscience applications. 

I thank Liang Fu and T. Senthil for numerous insightful discussions, and I acknowledge useful conversations with L. Levitov, P. Lee, K. Michaeli, S. Parameswaran, A.V.Chubukov and C. Varma.


\begin{thebibliography}{99}
\vspace{-7mm}

\providecommand{\natexlab}[1]{#1}
\providecommand{\url}[1]{\texttt{#1}}
\expandafter\ifx\csname urlstyle\endcsname\relax
  \providecommand{\doi}[1]{doi: #1}\else
  \providecommand{\doi}{doi: \begingroup \urlstyle{rm}\Url}\fi
  
  \bibitem[Volovik ()]{Volovikold}
%  G.E.Volovik and V.P.Mineev, %Textures, vortices, and superfluidity of $^3$He,  
%{\it Soviet Phys. Usp.} {\bf 25}, 187 (1982) 
Volovik, G.E. %Quantized Hall Effect in Superfluid Helium-3 Film. 
{\it Phys. Lett. {\bf A}} {\bf 128}, 277-279 (1988)

%\bibitem[Volovikold2 (1988)]{Volovikold2}
%Volovik, G.E. An analog of the quantum Hall effect in a superfluid ${}^3$He film {\it JETP} {\bf 67}, 1804-1811 (1988)

\bibitem[Sigrist (1991)]{Sigrist}
Sigrist, M. and Ueda, K. %Phenomenological theory of unconventional superconductivity. 
{\it Rev. Mod. Phys.} {\bf 63}, 239 (1991).

\bibitem[Sigrist (1991)]{Sigrist_1}
%Sigrist, M.  Time-Reversal Symmetry Breaking States in High-Temperature Superconductors. 
M. Sigrist, {\it Prog. Theor. Phys.} {\bf 99}, 899 (1998).

\bibitem[Sachdev (2000)]{Vojta} %Vojta M., Zhang Y.,  Sachdev S. Quantum phase transitions in d-wave superconductors.  
M. Vojta, Y. Zhang and S. Sachdev, {\it Phys. Rev. Lett.} {\bf 85}, 4940 (2000).

  \bibitem[Fu (2008)]{Kane}
%Fu, L. and Kane, C.L. Superconducting Proximity Effect and Majorana Fermions at the Surface of a Topological Insulator.
L. Fu and C.L.Kane, {\it Phys. Rev. Lett.} {\bf 100}, 096407 (2008)

\bibitem[Qi (2009)]{Qi}
% Qi X.L.,  Hughes T., Raghu S., and  Zhang S-C, Time-Reversal-Invariant Topological Superconductors and Superfluids in Two and Three Dimensions
X.L.Qi, T. Hughes, S. Raghu and S.C.Zhang, {\it Phys. Rev. Lett.} {\bf 102}, 187001 (2009)

\bibitem[Cheng (2010)]{Cheng}
%Cheng, M. Sun, K. Galitski, V. and Das Sarma, S. Stable topological superconductivity in a family of two-dimensional fermion models. 
M. Cheng, K. Sun, V. Galitski and S. Das Sarma, {\it Phys. Rev. B} 81, 024504 (2010)

\bibitem[Schnyder (2010)]{Schnyder}
Schnyder A., Ryu S., Furusaki A., and Ludwig A., 
% Classification of topological insulators and superconductors in three spatial dimensions. 
{\it Phys. Rev. B} {\bf 78}, 195125 (2010);


 
%\bibitem[Berz (2010)]{Berz} 
%Fu L.  and Berz E., Odd-parity topological superconductors: Theory and allication to Cu$_x$Bi$_2$Se$_3$. {\it Phys. Rev. Lett.}  {\bf 105}, 097001 (2010).

%\bibitem[Chung (2011)] {Chung} 
%Chung, S. B., Zhang, H.-J., Qi, X-L, , and  Zhang S-C.,
%Topological superconductivity and Majorana fermions in half-metal / superconductor heterostructure, arXiv:1011.6422.  

%\bibitem[DasSarma (2011)]{DasSarma}
%DasSarma, S., Adam, S., Hwang, E.H. and Rossi, E. {\it Rev. Mod. Phys.} {\bf 83}, 407-470 (2011)

%\bibitem[CastroNeto (2009)]{CastroNeto}
%Castro Neto, A.H., Guinea, F., Peres, N.M.R., Novoselov, K.S. and Geim, A.K. The electronic properties of graphene. {\it Rev. Mod. Phys.} {\bf 81}, 109-162 (2009)

\bibitem[Mackenzie (2003)]{Mackenzie}  
%Mackenzie, A.P. and  Maeno, Y., The superconductivity of Sr2RuO4 and the physics of spin-triplet pairing
A.P.Mackenzie and Y. Maeno, {\it Rev. Mod. Phys.}, {\bf 75}, 657 (2003).
 
 \bibitem[Nandkishore (2012)]{Nandkishore}
 %Nandkishore, R., Levitov, L.S. and Chubukov, A.V., Chiral superconductivity from repulsive interactions in doped graphene. 
 R. Nandkishore, L.S.Levitov and A.V.Chubukov, {\it Nature Physics}, 8, 158-163 (2012)
 
 \bibitem[Thomale (2012)]{Thomale}
%Kiesel, M., Platt, C., Hanke, W., Abanin, D.A. and Thomale, R. Competing many-body instabilities and unconventional superconductivity in graphene. 
M. Kiesel, C. Platt, W. Hanke, D.A.Abanin and R. Thomale, http://arxiv.org/pdf/1109.2953.pdf (2011)
 
 \bibitem[Kivelson (1994)]{Kivelson}
% Emery, V.J. and Kivelson, S.A., Importance of phase fluctuations in superconductors with small superfluid density, 
V.J.Emery and S.A.Kivelson, {\it Nature}, 374, 434 - 437 (1994).

\bibitem[Xu (2007)]{Xu}
C. Xu,% Phase transitions in coupled two dimensional XY systems with spatial anisotropy. 
http://arxiv.org/abs/0706.1609v2 (2007) (unpublished). 

\bibitem[Mineev (1998)]{Mineev}
V.P. Mineev and K.V. Samokhin, {\it Introduction to Unconventional Superconductivity}, {\it Gordon and Breach Science Publishers}, (1998), p69

\bibitem[JKNN (1977)]{JKNN}
J. V. JosŽ, L. P. Kadanoff, S. Kirkpatrick, and D. R. Nelson, Phys. Rev. B 16, 1217 (1977)
% Renormalization, vortices, and symmetry-breaking perturbations in the two-dimensional planar model

\bibitem[Sondhi (2005)]{Sondhi}
V. Oganesyan, D.A.Huse and S.L.Sondhi, Phys. Rev. B. 73, 094503, (2006)

\bibitem[Roy (2008)]{Roy}
R. Roy and C. Kallin, Phys. Rev. B 77, 174513 (2008)

\bibitem[Senthil (1998)]{Senthil}
T. Senthil, J.B.Marston and M.P.A.Fisher, Phys. Rev. B 60, 4245Ð4254 (1999)

\bibitem[Kerr (2011)]{Kerr}
%Nandkishore, R. and Levitov, L.S., Polar Kerr Efect and Time Reversal Symmetry Breaking in Bilayer Graphene,  
R. Nandkishore and L.S.Levitov, Phys. Rev. Lett. 107, 097402 (2011)

\bibitem[Argyres (1957)]{Argyres}
%Argyres, Petros N., Theory of the Faraday and Kerr effects in ferromagnetics.
P.N. Argyres,  {\it Physical Review} 97, 334-345, (1955). 
 
 %\bibitem[Xiao-Gang (2000)]{Xiao-Gang}
 %Wen, X.G. {\it Quantum Field Theory of Many-body Systems: From the Origin of Sound to an Origin of Light and Electrons}, 
 %X.G.Wen,  {\it Quantum Field Theory of Many-body Systems: From the Origin of Sound to an Origin of Light and Electrons}, Oxford Graduate Texts, Oxford, 2007. 
 
% \bibitem[Sachdev (2000)]{Sachdev}
%S. Sachdev. {\it Quantum Phase Transitions}. Cambridge University Press (2011)

%\bibitem[Volovik (1982]{Volovik} G.E.Volovik and V.P.Mineev, %Textures, vortices, and superfluidity of $^3$He,  
%{\it Soviet Phys. Usp.} {\bf 25}, 187 (1982) 

\bibitem[Kivelson2 (2000)]{Kivelson2}
%Jamei, R. Kivelson, S. and Spivak, B. Universal Aspects of Coulomb-Frustrated Phase Separation.
R. Jamei, S. Kivelson and B. Spivak, {\it Phys. Rev. Lett.}, 94, 056805 (2005)

%\bibitem[Hasan (2010)]{Hasan}
%Hasan, M.Z. and Kane, C.L. Topological insulators. {\it  Rev. Mod. Phys.} {\bf 82}, 3045 (2010)


%\bibitem[Goryo (1999)]{Goryo}
%Goryo, J. and Ishikawa, K. Observation of induced chern-simons term in P- and T- violating superconductors. {\it Physics Letter A}, {\bf 260}, 294-299 (1999)

%\bibitem{Horovitz}
%Horovitz, B. and Golub, A. Superconductors with broken time-reversal symmetry: Spontaneous magnetization and quantum Hall effects. {\it Phys. Rev. B} {\bf 68}, 214503 (2003)

%\bibitem[Kapitulnik (1998)]{kapitulnik} 
%Laughlin, R.B. Magnetic Induction of $d_{x^2-y^2}+id_{xy}$ Order in High-$T_c$ Superconductors. {\it Phys. Rev. Lett.} {\bf 80}, 5188 (1998) 

%\bibitem[Sato (2010)]{Sato}
%Sato, M., Takahashi, Y. and Fujimoto, S. Non-Abelian topological orders and Majorana fermions in spin-singlet superconductors. {\it Phys. Rev. B} {\bf 82}, 134521 (2010)

%\bibitem[Mao (2011)]{Mao}
 %Mao L.,  Shi J., Niu Q., and Zhang C. Superconducting phase with a chiral $f$-wave pairing symmetry and Majorana fermions induced in a hole-doped 
%semiconductor.  {\it Phys. Rev. Lett} {\bf 106}, 157003 (2011). 


%\bibitem[Jiang (2008)]{Jiang}
%Jiang, Y., Yao, D.X., Carlson, E.W., Chen, H.D. and Hu, J.P. Andreev conductance in the $d+id$-wave superconducting state of graphene. {\it Phys. Rev. B} {\bf 77}, 235420 (2008)

\bibitem[Xia (2006)]{Xia}
J. Xia, Y. Maeno, P. T. Beyersdorf, M. M. Fejer and A.Kapitulnik, Phys. Rev. Lett. 97, 167002 (2006).
%\bibitem[Tewari (2008)]{Tewari}
%Tewari, S., Zhang, C., Yakovenko, V.M. and DasSarma, S. Time-Reversal Symmetry Breaking by a ($d+id$) Density-Wave state in Underdoped Cuprate Superconductors. {\it Phys. Rev. Lett.} {\bf 100}, 217004 (2008)

%\bibitem[Dagan (2002)]{Dagan}
%Dagan, Y. and Deutscher, G. On the origin of time reversal symmetry breaking in $Y_{1-y} Ca_y Ba_2 Cu_3 O_{7-x}$. {\it Europhys. Lett.} {\bf 57}, 444 (2002).

%\bibitem[Kivelson (2)]{Kivelson}
%Carlson, E.W., Emery, V.J., Kivelson, S.A. and Orgad, D. Concepts in High Temperature Superconductivity. {\it Superconductivity. Conventional and Unconventional Superconductors.} Springer-Verlag (Berlin) (2008)


%\bibitem[Gonzalez (2008)]{Gonzalez}
%Gonzalez, J. Kohn Luttinger superconductivity in Graphene. {\it Phys. Rev. B} {\bf 78}, 205431 (2008)
%\bibitem[Li (2004)]{Li}
%Li, J.X. and Wang, Z.D. Spin-resonance peak in $Na_x Co O_{2-x} H_2O $ superconductors: A probe of the pairing symmetry. {\it Phys. Rev. B} {\bf 70}, 212512 (2004)

%\bibitem[Uchoa (2007)]{Uchoa}
%Uchoa B. and Castro Neto, A.H. Superconducting States of Pure and Doped Graphene. {\it Phys. Rev. Lett.} {\bf 98}, 146801 (2007) 
%AC \addRN{and citing articles}
% and references therein.

%\bibitem[Raghu (2011)]{Raghu}
%Raghu, S. and Kivelson, S.A. Superconductivity from repulsive interactions in the two dimensional electron gas. {\it Phys. Rev. B} 83, 094518 (2011)

%\bibitem[Shankar (1994)]{Shankar}
%Shankar, R. Renormalization group approach to interacting fermions. {\it Rev. Mod. Phys.} {\bf 66}, 129 (1994)

%\bibitem[McChesney (2010)]{McChesney}
%McChesney, J.L., Bostwick, A., Ohta, T., Seyller, T., Horn, K., Gonzalez, J. and Rotenberg, E. Extended van Hove Singularity and Superconducting Instability in Doped Graphene. {\it Phys. Rev. Lett.} {\bf 104}, 136803 (2010)

%\bibitem[Kopnin (2008)]{Kopnin}
%Kopnin, N.B. and Sonin, E.B. BCS Superconductivity of Dirac Electrons in Graphene Layers. {\it Phys. Rev. Lett.} {\bf 100}, 246808 (2008)

%\bibitem[Lozovik (2010)]{Lozovik}
 %Lozovik, Y.E., Ogarkov, S.L. and Sokolik, A.A. Theory of superconductivity for Dirac electrons in graphene. {\it JETP}, {\bf 110}, 1, pp 49-57 (2010)


%\bibitem[Gonzalez (1999)]{Guinea}
%Gonzalez, J., Guinea F., and Vozmediano, M.A.H. Marginal-Fermi-liquid behavior from two-dimensional Coulomb interaction. {\it Phys. Rev. B} {\bf 59}, 2474(R) (1999)

%\bibitem[Wallace (1947)]{Wallace}
%Wallace, P.R. The Band Theory of Graphite. {\it Phys. Rev.} {\bf 71}, 622-634 (1947)

%\bibitem[Schulz (1987)]{Shulz}
%Schulz, H.J. Superconductivity and Antiferromagnetism in the Two-Dimensional Hubbard Model: Scaling Theory. {\it Europhys. Lett.} {\bf 4} 609 (1987) 

%\bibitem[Dzyaloshinskii (1987)]{Dzyaloshinskii}
%Dzyaloshinskii, I.E. Maximal inrease of the superconducting transition temperature due to the presence or van't Hoff singularities. {\it Sov. Phys. JETP} {\bf 66}, 848 (1987)

%\bibitem[Furukawa (1998)]{Furukawa}
%Furukawa, N., Rice, T.M. and Salmhofer, M. Truncation of a Two-Dimensional Fermi Surface due to Quasiparticle Gap formation at the Saddle Points. {\it Phys. Rev. Lett.} {\bf 81}, 3195 (1998)

%\bibitem[Rice (2009)]{ricereview} 
%LeHur, K. and Rice, T.M. Superconductivity close to the Mott state: From condensed-matter systems to superfluidity in optical lattices. {\it Annals of Physics} {\bf 324} (2009) 1452-1515

%\bibitem[Batista (2008)]{Batista} Martin, I., and Batista, C.D., Itinerant electron-driven chiral magnetic ordering and spontaneous quantum hall effect in triangular lattice models. {\it Phys. Rev. Lett.} {\bf 101}, 156402 (2008) 

%\bibitem[Chubukov (2010)]{chubukov} Maiti, S. and Chubukov, A.V. Renormalization group flow, competing phases and the structure of superconducting gap in multiband models of iron-based superconductors. {\it Phys. Rev. B} {\bf 82}, 214515 (2010)

%\bibitem[Thomale (2011)]{Thomale}
%Thomale, R., Platt, C., Hanke, W. and Bernevig, B.A. Mechanism for Explaining Differences in the Order Parameters of $FeAs$-Based and $FeP$-Based Pnictide Superconductors. {\it Phys. Rev. Lett.} {\bf 106}, 187003 (2011)

%\bibitem[Vozmediano (2011)]{Vozmediano}
%Valenzuela B. and Vozmediano, M.AH. Pomeranchuk instability in doped graphene. {\it New. J. Phys.}  {\bf 10} (2008) 113009 

%\bibitem[TaoLi (2011)]{TaoLi}
%Li,T. Spontaneous quantum Hall effect in quarter doped Hubbard model on honeycomb lattice and its possible realization in quarter doped graphene system. {\it arxiv}: 1103.2420 (2011) (unpublished)

%\bibitem[MoraisSmith (2011)]{MoraisSmith}
%Makogon, D., van Gelderen, R., Roldan, R. and Morais Smith, C. Spin-density-wave instability in graphene doped near the van Hove singularity. {\it arxiv}: 1104.5334 (2011) (unpublished)


%\bibitem[Son (1999)]{Son}
%Son, D.T. Superconductivity by long-range color magnetic interaction in high-density quark matter. {\it Phys. Rev. D} {\bf 59}, 094019 (1999)

%\bibitem[qc3D (2011)]{qcp} 
%see e.g., Moon, E.G. and Chubukov, A.V. Quantum-critical pairing with varying exponents. {\it arXiv} :1005.0356 (unpublished) and references therein.

%\bibitem[Baskaran (2008)]{Baskaran}
%Pathak, S., Shenoy, V.B. and Baskaran, G. Possibility of High $T_c$ Superconductivity in doped Graphene. {\it arXiv}: 0809.0244 (2008) (unpublished)

%\bibitem[Doniach (2009)]{Doniach}
%Black-Schaffer, A.M. and Doniach, S. Resonating valence bonds and mean-field $d$-wave superconductivity in graphite. {\it Phys. Rev. B} {\bf 75}, 134512 (2007)

%\bibitem[Honerkamp (2008)]{Honerkamp}
%Honerkamp, C.,  Density Waves and Cooper Pairing on the Honeycomb Lattice. {\it Phys. Rev. Lett.} {\bf 100}, 146404 (2008)

%\bibitem[Roy (2010)]{Roy}
%Roy, B. and Herbut, I.F. Unconventional superconductivity on honeycomb lattice: Theory of Kekule order parameter. {\it Phys. Rev. B} {\bf 82}, 035429 (2010)

% Roy, R., and Kallin, C., Collective modes and electromagnetic response of a chiral superconductor. {\it Phys. Rev. B} 77, 174513 (2008)

\bibitem[Ong (2010)]{Ong}
L. Li, N. Alidoust, J.M.Tranquada, G.D.Gu and N.P.Ong, Phys. Rev. Lett. 107, 277001 (2011)

%\bibitem[supplement (2011)]{supplement}
%See online supplementary material. 



\end{thebibliography}
\end{document}